\documentclass[10pt]{article}
\usepackage{amsmath}
\usepackage{amssymb}
\usepackage{mathrsfs}
\usepackage{cases}
\usepackage{graphicx, subfigure,}
\usepackage{caption}
\usepackage{graphicx}
\renewcommand{\paragraph}{\roman{paragraph}}

\def \lamax{\lambda_{\max}}
\def \sigmax{\sigma_{\max}}

\def \lamin{\lambda_{\min}}
\def \sigmin{\sigma_{\min}}

\def \e{\epsilon}

\def \e{\epsilon}

\newtheorem{theorem}{\scshape \mdseries  Theorem}[section]
\newtheorem{lemma}[theorem]{\scshape \mdseries  Lemma}

\begin{document}

\title{\sf Mixed Compressed Sensing Based on Random Graphs\thanks{
Supported by National Natural Science Foundation of China (11071002),
Program for New Century Excellent Talents in University (NCET-10-0001),
Key Project of Chinese Ministry of Education (210091),
Specialized Research Fund for the Doctoral Program of Higher Education (20103401110002),
Science and Technological Fund of Anhui Province for Outstanding Youth (10040606Y33),
Scientific Research Fund for Fostering Distinguished Young Scholars of Anhui
University(KJJQ1001), Academic Innovation Team of Anhui University Project (KJTD001B).}}
\author{Yi-Zheng Fan\thanks{Corresponding author.
 E-mail addresses: fanyz@ahu.edu.cn(Y.-Z. Fan), huangtaooo@hotmail.com (T. Huang), zhu\_m@163.com (M. Zhu)}, Tao Huang, Ming Zhu\\
  {\small  \it $1.$ Key Laboratory of Intelligent Computing and Signal Processing of Ministry of Education,}\\
{\small \it Anhui University, Hefei 230039, P. R. China} \\
  {\small  \it $2.$ School of Mathematical Sciences, Anhui University, Hefei 230601, P. R. China} \\
 }
\date{}
\maketitle

\noindent
{\bf Abstract}\  \
Finding a suitable measurement matrix is an important topic in compressed sensing.
Though the known random matrix, whose entries are drawn independently from a certain probability distribution, can be used
  as a measurement matrix and recover signal well, in most cases, we hope the measurement matrix imposed with some special structure.
  In this paper, based on random graph models, we show that the mixed symmetric random matrices,
  whose diagonal entries obey a distribution and non-diagonal entries obey another distribution,
     can be used to recover signal successfully with high probability.



\newpage
\indent

\section{Introduction}

The {\it Compressed Sensing } problem is: recovering $x$ from knowledge of $y=\Phi x$ where $\Phi$ is a suitable  $n \times N$ measurement matrix and $n < N$. This problem has a number of potential applications in signal processing, as well as other areas of science and technology. In 2006, the area of compressed sensing made great progress by two ground breaking papers, namely \cite{donoho4} by Donoho and  \cite{candes} by Cand\`es, Romberg and Tao. From then on, plenty of theoretical papers of compressed sensing are published.

It's well known now that recovering $x$ can be solved by $l_1$-minimization instead of $l_0$-minimization:
$$\min\|x\|_1     \mbox{~~subject to~~}    y=\Phi x\eqno(1.1)$$
where the $l_p$-norm is defined  $\|x\|_p=(\sum_{j=1}^n |x_j|^p)^{1/p}$, as usual.

Without loss of generality, we assume that the arbitrary vector $x=(x_i)_{i=1}^n \in \mathbb{R}^n$ is {\em $k$-sparse}, if  the number of non-zero coefficients of vector $x$ is at most $k$. Measurement matrices are required to satisfy certain conditions such as, for instance, {\em Restricted Isometry Property} (abbreviated as RIP) \cite{can}.
An $n \times N$ matrix $\Phi$ is said to have RIP of order $k$ if there exists  $\delta_k \in (0,1)$ such that
$$(1-\delta_k)\|x\|_2^2 \le \|\Phi x\|_2^2 \le (1+\delta_k)\|x\|_2^2 \eqno(1.2)$$
for all $k$-sparse vectors $x$.

The problem, how to choose a suitable measurement matrix $\Phi$, must be investigated in this field.
Most of them are random matrices such as Gaussian or Bernoulli random matrices as well as partial Fourier
matrices\cite{candes3,rau3,rude,bar,candes}, if we are allowed to choose the sensing matrix freely.
Although Gaussian random matrix and others are optimal for sparse recovery, they have limited using in practice because many measurement technologies impose structure on the matrix.

It is still an open question whether deterministic matrices
can be carefully constructed to have similar properties with respect to compressed sensing
problems.  Actually, most applications do not allow a free choice of the sensing matrix and enforce a particularly
structured matrix. Recently, Bajwa {\it et al.} estimated a random Toeplitz type or circulant matrix, where the entries of the vector generating the Toeplitz or circulant matrices are chosen at random according to a suitable probability distribution, which then allowed for providing recovery guarantees for $l_1$-minimization; see \cite{raz,hol,pfa,wakin,raz2}.
Compared to Bernoulli or Gaussian matrices, random Toepliz and circulant matrices have the advantages that they require a reduced number of random entries to be generated.
They close the theoretical gap by providing recovery guarantees for $l_1$-minimization in connection with circulant or Toeplitz type matrices where the necessary number of measurements scales linearly with the sparsity.

In this paper we focus our attention on the matrices associated with random graphs.
The classical {\em Erd\"os-R\'enyi model} $\mathscr{G}_N(p)$ consists of all graphs on $N$ vertices in which the edges (including loops) are chosen
independently with probability $p\in (0,1)$ (see \cite{bollo}), where a loop is an edge joining one vertex to itself.
If letting $A(G)=(a_{ij})$ be the adjacency matrix of a graph $G \in \mathscr{G}_N(1/2)$, that is, $a_{ij}=1$ if $ij$ is an edge of $G$ and $a_{ij}=0$ otherwise,
then $2A(G)-J$ is a random symmetric matrix whose entries hold Bernoulli distribution, where $J$ is a matrix consisting of all ones.

The above graph $G$ can be viewed as a weighted graph with each edge having weight $1$.
In formal, a {\em weighted graph} $G(\omega)$ is one with each edge $ij$ assigned a nonzero weight $\omega_{ij}$.
If there is no edge between vertices $i$ and $j$, we may think that the weight of $ij$ is zero, or $\omega_{ij}=0$.
The adjacency matrix of $G(\omega)$ is $A(G(\omega))=(\omega_{ij})$.

We now consider the {\em mixed weighted random graph model} $\mathscr{G}_N(\mathbf{P}_1,\mathbf{P}_2)$, which consists of all graphs on $N$ vertices
for which the weights $\omega_{ii}$ are chosen independently that obey the probability distribution $\mathbf{P}_1$, and $\omega_{ij}\,(i \ne j)$ are chosen
independently that obey the probability distribution $\mathbf{P}_2$.
For example, $\mathbf{P}_1$ or $\mathbf{P}_2$ is the Bernoulli distribution, or normal Gaussian distribution, or is the following {\it 3-point distribution}:
$$ x= \left\{
\begin{array}{cc}
+\sqrt{3}       & \hbox{with probability~~~} 1/6,\\
0        & \hbox{with probability~~~} 2/3,\\
-\sqrt{3}       & \hbox{with probability~~~} 1/6.
\end{array}
\right. \eqno(1.1)
$$
In this case, the matrix $A(G)$ is a mixed symmetric random matrix based on the graph structure of $G$.
In the Fig 1.1, a random graph is generated, where $\omega_{ii}$'s obey the normal Gaussian distribution
and $\omega_{ij}$'s $(i \ne j)$ obey the Bernoulli distribution.

     \begin{figure}[!h]
  \centering
  \renewcommand\thefigure{\arabic{section}.\arabic{figure}}
    \begin{minipage}[]{.6\textwidth}
      \centering
     \includegraphics[width=1\textwidth]{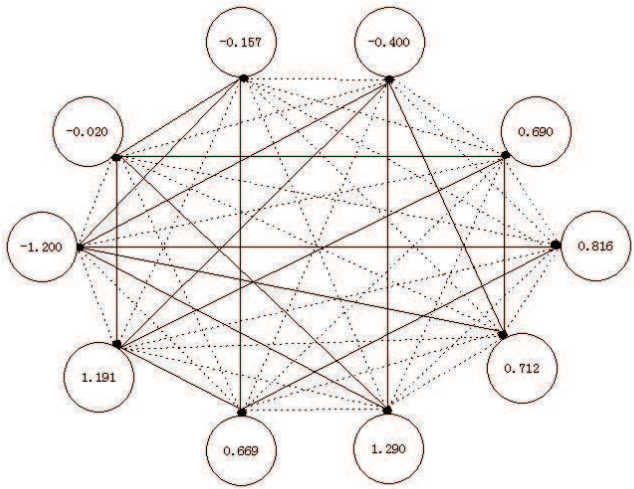}\\
    \caption{ A random graph with black lines having weight $1$ and dotted lines having weight $-1$ }
   \end{minipage}
     \hspace{0in}
\end{figure}

In the previous work \cite{Fan} we have shown that random symmetric Bernoulli matrix satisfied RIP well and can be used as measurement matrix either.
In this paper we discuses restricted isometry for symmetric random matrices based on random graphs in a more general setting.

\section{Preliminaries}

We list some important lemmas for us.
Denote by $\sigmin(A),\sigmax(A)$ the minimum and maximum singular values of the matrix $A$,
and by $\lamax(B),\lamin(B)$ the minimum and maximum eigenvalues of the symmetric matrix $B$.

\begin{lemma} {\em\cite{bai1,yin}}
Let $A$ be a $n \times p$ matrix whose entries are iid random variables with zero mean, unit variance and finite fourth moment.
If $n \to \infty, p \to \infty$ and $ p/n \to y \in (0,1)$, then
$$ \sigmin(n^{-1/2}A) \to 1-\sqrt{y} \mbox{~a.s.}, ~ \sigmax(n^{-1/2}A) \to 1+\sqrt{y}  \mbox{~a.s.}$$
\end{lemma}

\begin{lemma}{\em \cite{bai}}
Let $A$ be a $n \times n$ random symmetric matrix whose entries above the diagonal are iid random variables with variance $\sigma^2$,
the entries on the diagonal are iid random variables, and the diagonal entries are independent of nondiagonal entries.
If $E((A_{11}^+)^2)<\infty$, $E(A_{12}^4)<\infty$, $E(A_{12}) \le 0$,
then $$\lamax(n^{-1/2}A) \to 2\sigma \mbox{~a.s.}$$
\end{lemma}

\begin{lemma}{\em \cite{can}}
Assume that $\delta_{2k}<\sqrt{2}-1$. Then the solution $x^*$ to (1.2) obeys
$$\|x^*-x\|_1\le C_0\|x-x_{(k)}\|_1, \; \|x^*-x\|_2 \le C_0 k^{-1/2} \|x-x_{(k)}\|_1,$$
for some constant $C_0$, where $x_{(k)}$ is obtained from $x$ by setting all but the $k$-largest entries to be zero.
In particular if $x$ is $k$-sparse, the recovery is exact.
\end{lemma}

If the measurements are corrupted with noise, that is $y = \Phi x +z$,
where $z$ is an unknown noise term.
we will consider the following problem:
$$\min_{x \in \mathbb{R}^N}\|x\| \mbox{~~subject to~~} \|y-\Phi x\|\le \e,\eqno(2.1)$$
where $\e$ is an upper bound on the size of the noisy contribution.

\begin{lemma}{\em \cite{can}}
Assume that $\delta_{2k}<\sqrt{2}-1$ and $\|z\|_2 \le \e$. Then the solution $x^*$ to (2.1) obeys
$$\|x^*-x\|_2 \le C_0 k^{-1/2} \|x-x_k\|_1+C_1\e,$$
for some constants $C_0,C_1$.
\end{lemma}

\section{Main results}

Now let $G(\omega) \in \mathscr{G}_N(\mathbf{P}_1,\mathbf{P}_2)$, where
 $$ E((\omega_{11}^+)^2)< \infty, E(\omega_{12})=0, E(\omega_{12}^2)=\sigma^2, E(\omega_{12}^4)< \infty. \eqno(3.1)$$
Let $\Psi$ be obtained from the adjacency matrix $A(G(\omega))$ by arbitrarily choosing a subset $\Theta$ of $n$ rows, and $\Phi=n^{-1/2}\Psi$.
W.l.g, take $\Theta=\{1,2,\ldots,n\}$.

\begin{theorem}
Let $G(\omega)$ be a random graph holding (3.1), and
let $\Psi$ be obtained from the adjacency matrix $A(G(\omega))$ by arbitrarily choosing $n$ rows,
and let $\Phi=n^{-1/2}\Psi$.
For any given $\delta_k$ such that $0<\delta_k<1$, letting $\frac{k}{n} \to \gamma$, if $\gamma$ is close to $0$, there
always exists some $\sigma$ such that the matrix $\Phi$ holds RIP a.s.
\end{theorem}

{\bf Proof:}
Let $ S \subset \{1,2,\ldots,N \}$ be a index subset of cardinality $k$, and let
 $\Phi_S$ be the submatrix of $\Phi$ consisting of the columns indexed by $S$.
RIP implies that
$$ 1-\delta_k \le \lamin(\Phi_S^T\Phi_S) \le \lamax(\Phi_S^T\Phi_S) \le 1+\delta_k.\eqno(3.2)$$
Suppose $\frac{k}{n} \to \gamma$ in the following.

{\it Case 1:} $ S \subset \Theta$. W.l.g, take $S=\{1,2,\ldots,k\}$.
Write $\Phi_S=\left[\begin{array}{c}A \\B \end{array}\right]$, where $A$ is a random symmetric matrix of order $k$ and $B$ is a random matrix of size $(n-k)\times k$ with i.i.d elements.
Then
$$ \Phi_S^T\Phi_S=\left[\begin{array}{cc}A^T & B^T \end{array}\right] \left[\begin{array}{c}A \\B \end{array}\right]
=A^TA+B^TB.$$
Since $A^TA$ is positive semidefinite,
$$  \lamin(B^TB) \le \lamin(\Phi_S^T\Phi_S) \le \lamax(\Phi_S^T\Phi_S) \le \lamax(A^TA)+\lamax(B^TB).$$
By Lemma 2.1
\begin{align*}
\lamin(B^TB) & \to \lim_{n \to \infty} \frac{n-k}{n}\left(1-\sqrt{\frac{k}{n-k}}\right)^2\sigma^2=(1-\gamma)\left(1-\sqrt{\frac{\gamma}{1-\gamma}}\right)^2 \sigma^2\\
& =\left(1-2\sqrt{\gamma(1-\gamma)}\right)\sigma^2.~~~~~ \hbox{a.s.}
\end{align*}
Similarly, by Lemma 2.1 and Lemma 2.2,
$$\lamax(A^TA)+\lamax(B^TB) \to \left(1+4\gamma+2\sqrt{\gamma(1-\gamma)}\right)\sigma^2.~~~~~ \hbox{a.s.}$$
Hence, if taking $\sigma$ such that
$$ \frac{1-\delta_k}{1-2\sqrt{\gamma(1-\gamma)}} \le \sigma^2 \le \frac{1+\delta_k}{1+4\gamma+2\sqrt{\gamma(1-\gamma)}},  \eqno(3.3)$$
then the inequalities (3.2) hold.

{\it Case 2:} $ S \subset \{1,2,\ldots,N \}\setminus \Theta$.
Then $\Phi_S$ is a random matrix of size $n \times k$ with i.i.d elements.
By Lemma 2.1,
$$\lamin(\Phi_S^T\Phi_S) \to (1-\sqrt{\gamma})^2 \sigma^2, ~~~ \lamax(\Phi_S^T\Phi_S) \to (1+\sqrt{\gamma})^2 \sigma^2.$$
Now taking $\sigma$ such that
$$ \frac{1-\delta_k}{(1-\sqrt{\gamma})^2} \le \sigma^2 \le \frac{1+\delta_k}{(1+\sqrt{\gamma})^2},\eqno(3.4)$$
 then the inequalities (3.2) hold.

{\it Case 3:} $ S \cap \Theta \ne \emptyset$ and $S \cap (\{1,2,\ldots,N \}\setminus \Theta)\ne \emptyset $.
 W.l.g, take $S\cap \Theta=\{1,2,\ldots,k_1\}$, where $1 \le k_1 < k$.
 Write $\Phi_S=\left[\begin{array}{c}A \\B \end{array}\right]$, where $A=[A_1,A_2]$ and $A_1$ is a random symmetric matrix of order $k_1$ and $A_2$ is a random matrix of size $k_1\times (k-k_1)$ with i.i.d elements, $B$ is a random matrix of size $(n-k_1)\times k$ with i.i.d elements.
By a similar discussion as in Case 1,
\begin{align*} \lamin(\Phi_S^T\Phi_S) & \ge \lamin(B^TB) \to \lim_{n \to \infty} \left(1-\sqrt{\frac{k}{n-k_1}}\right)^2 \sigma^2 \ge  \lim_{n \to \infty} \left(1-\sqrt{\frac{k}{n-k}}\right)^2 \sigma^2 \\
&=\left(1-\sqrt{\frac{\gamma}{1-\gamma}}\right)^2 \sigma^2~~~~~ \hbox{a.s.}
\end{align*}
Assuming $k_1 \ge k/2$, we also have
\begin{align*}
\lamax(\Phi_S^T\Phi_S) & \le \lamax(A^TA)+\lamax(B^TB) \\
& \le  \lamax(A_1^TA_1)+\lamax(A_2^TA_2)+\lamax(B^TB)\\
& \to \lim_{n \to \infty} \frac{k_1}{n}(2 \sigma)^2+ \frac{k_1}{n} \left(1+\sqrt{\frac{k-k_1}{k_1}}\right)^2\sigma^2+
\frac{n-k_1}{n} \left(1+\sqrt{\frac{k}{n-k_1}}\right)^2\sigma^2\\
&\le (1+7 \gamma +2 \sqrt{\gamma})\sigma^2 ~~~~~ \hbox{a.s.}
\end{align*}
So, if taking $\sigma$ such that
$$ \frac{1-\delta_k}{\left(1-\sqrt{\frac{\gamma}{1-\gamma}}\right)^2} \le \sigma^2 \le \frac{1+\delta_k}{(1+7 \gamma +2 \sqrt{\gamma})},\eqno(3.5)$$
 then the inequalities (3.2) hold.
The other case of $k_1 < k/2$ can be discussed similarly and is omitted.

In the inequalities (3.3)-(3.5), if taking $\gamma=0$, then $1-\delta_k \le \sigma^2 \le 1+\delta_k$.
As both sides of each of (3.3)-(3.5) are continuous in $\gamma$, if $\gamma \to 0$, then the inequalities (3.3)-(3.5) still hold.
So, for any given $\delta_k$ such that $0<\delta_k<1$, if $\gamma$ is sufficiently close to $0$, there
always exists some $\sigma$ such that the matrix $\Phi$ holds RIP a.s. \hfill $\blacksquare$

{\bf Remark:}
(1) For the random graph $G(\omega) \in \mathscr{G}_N(\mathbf{P}_1,\mathbf{P}_2)$, $\mathbf{P}_1$ or $\mathbf{P}_2$ can be taken as
the Bernoulli distribution, norm Gaussian distribution or the 3-point distribution as in (1.1), all of which satisfy (3.1) where $\sigma^2=1$.
So, we can construct a symmetric compressed sensing matrix, where the diagonal and nondiagonal entries obey different distributions.

(2) Here the result in Theorem 3.1 holds almost surely as $n$ goes to infinity.
As discussed in \cite{Fan}, if $\mathbf{P}_1=\mathbf{P}_2$ is the Bernoulli distribution,
we have a more accurate result in the below.

\begin{theorem}\cite{Fan}
For any given $0<\delta_k<1$, if taking $\Phi(\omega)=n^{-1/2} R$, and taking $n \ge c_1^{-1}k \log(N/k)$, then
RIP (1.2) holds for $\Phi(\omega)$ with the prescribed $\delta_k$ and order $k$ with probability $\ge 1-2e^{-c_2n}$,
where $c_1,c_2$ depend only on $\delta_k$.
\end{theorem}

\section{Experiment}

Let $x$ be a $k$-sparse discrete signal with length $256$ whose nonzero entries are $1$ or $-1$.
The classical convex optimization algorithm $\ell_1$-minimization is used for reconstruction.
The experimental results are compared with Gaussian,  Bernulli and symmetric mixed matrices (simply as S-Mixed).

   We first analysis the performances of these matrices under different sparsity. Set the  measurement number $n=100$.
    The results of $1000$ experiments are summarized in Fig. 4.1, from which
    we see that  all the performances  decrease while the sparsity increases,
    It is hard to distinguish which one is the best among these matrices.

     \begin{figure}[!h]
  \centering
  \renewcommand\thefigure{\arabic{section}.\arabic{figure}}
    \begin{minipage}[]{1\textwidth}
      \centering
     \includegraphics[width=.75\textwidth]{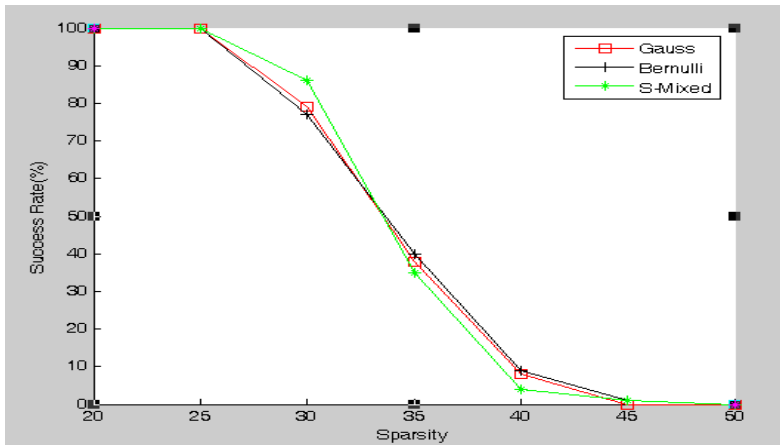}\\
    \caption{ Success rate as a function of sparsity K}
   \end{minipage}
     \hspace{0in}
\end{figure}

Next, We  investigate the performances of these matrices under different measurement numbers.
   Set the sparsity $k=20$.
   The results of $1000$ experiments are summarized and shown in Fig. 4.2.
 The performance of all matrices get better with the measurement number $n$ increasing.
   Especially, when $n \ge 95$ almost all experiments are successful.
 \begin{figure}[!h]
  \centering
  \renewcommand\thefigure{\arabic{section}.\arabic{figure}}
    \begin{minipage}[]{1\textwidth}
      \centering
     \includegraphics[width=.75\textwidth]{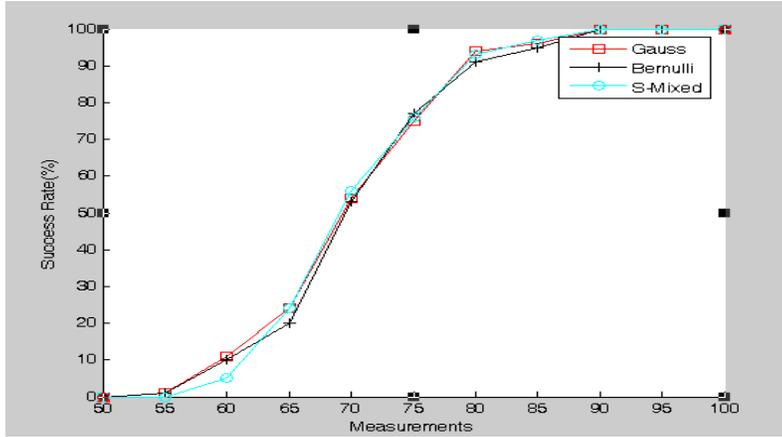}\\
    \caption{Success rate as a function of measurement number}
   \end{minipage}
     \hspace{0in}
\end{figure}
Now, we check the performances of the above measurement matrices through the real image reconstruction experiment.
   The original image is shown in Fig. 4.3, with size of $64\times 64$ and sparsity $k=739$.
   Set measurement number $n=2400$.
   The mean square error (MSE) is defined as $MSE=\frac{\|X-M\|_F}{\|M\|_F}$,
   where $\|\cdot\|_F$ being the Frobenius norm, $X$ is the reconstruction and $M$ is the original image.
The experimental results are shown in Fig. 4.3.
\begin{figure}[!h]
  \centering
  \renewcommand\thefigure{\arabic{section}.\arabic{figure}}
  \subfigure{
    \begin{minipage}{0.4\textwidth}
      \centering
     \includegraphics[width=0.8\textwidth]{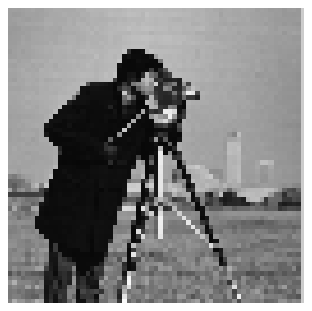}\\
   \caption*{Original image}
  \end{minipage}}
     \hspace{0in}
       \subfigure{
   \begin{minipage}{0.4\textwidth}
      \centering
     \includegraphics[width=0.8\textwidth]{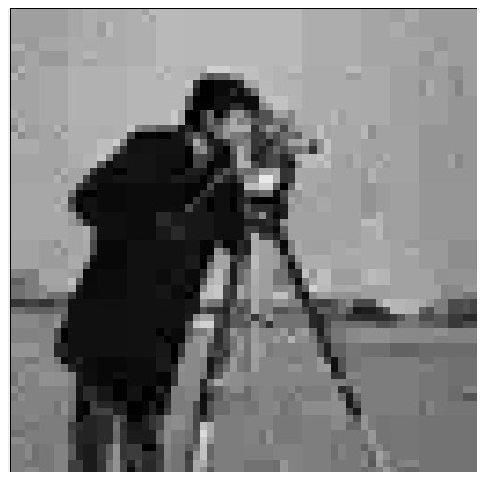}\\
    \caption*{S-Mixed(MSE=0.0673)}
   \end{minipage}}
   \hspace{0in}
    \subfigure{
   \begin{minipage}{0.4\textwidth}
      \centering
     \includegraphics[width=0.8\textwidth]{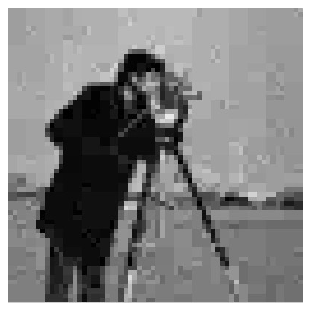}\\
    \caption*{Bernoulli(MSE=0.0672)}
   \end{minipage}}
   \hspace{0in}
  \subfigure{
    \begin{minipage}{0.4\textwidth}
      \centering
     \includegraphics[width=0.8\textwidth]{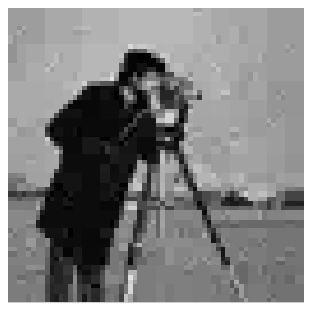}\\
    \caption*{Gaussian (MSE=0.0681)}
   \end{minipage}}
     \hspace{0in}
       \caption{ Real world data reconstruction}
\end{figure}
The experimental results show that these proposed matrices are suitable measurement matrices.

\small

\indent

\end{document}